\documentclass[12pt]{iopart}

\usepackage{lineno,hyperref}
\usepackage{ragged2e}
\usepackage{url}
\usepackage{graphics}
\usepackage{graphicx}
\usepackage{multirow}
\usepackage{array}
\usepackage{setspace}

\providecommand{\keywords}[1]{\small{\textit{Keywords:}} #1}

\begin{document}
\title{Tensor-driven extraction of developmental features from varying paediatric EEG datasets}

\author{E. Kinney-Lang$^{1,2,*}$, L. Spyrou$^1$, A. Ebied$^1$, R. FM Chin $^{2,3}$, and J. Escudero$^{1,2}$}
\address{$^1$School of Engineering, Institute for Digital Communications, The University of Edinburgh, Edinburgh EH9 3FB, United Kingdom}
\address{$^2$The Muir Maxwell Epilepsy Centre, The University of Edinburgh, Edinburgh EH8 9XD, United Kingdom}
\address{$^3$Royal Hospital for Sick Children, Edinburgh EH9 1LF, United Kingdom}
\address{$^*$Corresponding Author}
\ead{e.kinney-lang@ed.ac.uk}

\begin{abstract}
	\textit{Objective}. Consistently changing physiological properties in developing children's brains challenges new data heavy technologies, like brain-computer interfaces (BCI). Advancing signal processing methods in such technologies to be more sensitive to developmental changes could help improve their function and usability in paediatric populations. Taking advantage of the multi-dimensional structure of EEG data through tensor analysis offers a framework to extract relevant developmental features present in paediatric resting-state EEG datasets. 
	\textit{Methods}. Three paediatric datasets from varying developmental states and populations were analyzed using a developed two-step constrained Parallel Factor (PARAFAC) tensor decomposition. The datasets included the Muir Maxwell Epilepsy Centre, Children's Hospital Boston-MIT and the Child Mind Institute, outlining two impaired and one healthy population, respectively. Within dataset cross-validation used support vector machines (SVM) for classification of out-of-fold data predicting subject age as a proxy measure of development. t-distributed Stochastic Neighbour Embedding (t-SNE) maps complemented classification analysis through visualization of the high-dimensional feature structures.
	\textit{Main Results} Development-sensitive features were successfully identified for the developmental conditions of each dataset. SVM classification accuracy and misclassification costs were improved significantly for both healthy and impaired paediatric populations. t-SNE maps revealed suitable tensor factorization was key in extracting developmental features.
	\textit{Significance} The described methods are a promising tool for incorporating the unique developmental features present throughout childhood EEG into new technologies like BCI and its applications.
\end{abstract}

\maketitle
\section*{Introduction}
Diseases and injuries sustained in childhood are a major public health issue worldwide. The resultant acute and/or chronic motor disability affect millions, with early-life sustained motor insult potentially leading to learned non-use of afflicted regions and potential complications later in life \cite{Manning2015a,Johnston2004b}. While traditional therapeutic options for improving motor disability often involve exercise-based techniques \cite{Fluck2006a,Damiano2006} like constraint-induced movement therapy (CIM) \cite{Manning2015a,Taub2002,Damiano2010}, these exercise-based techniques have a trade-off in requiring residual movement and control in the patient's afflicted appendage \cite{Manning2015a,Millan2010b}. Brain-computer interfaces (BCI) and other emerging technologies are strong candidates for non-muscular neurorehabilitation options in clinical settings \cite{Ang2014,Mrachacz-Kersting2015,Wolpaw2007a,VanDokkum2015a}, with promising early results in adults \cite{Young2014,Soekadar2015}.

Accounting for varying electrophysiological properties in developing children, however, poses a hurdle for many of these data driven technologies, including BCI. Blending together engineering and medicine, BCIs provide direct communication channels between the brain and an output device, i.e. computer \cite{Wolpaw2002}, through advanced signal processing. Many popular BCI applications measure and decode electric potentials created in the brain, such as specific thought patterns used to invoke motor imagery (MI) \cite{Nicolas-Alonso2012,Ahn2014,Vuckovic2014}. Hardware like electroencephalography (EEG) record the electric potentials over the scalp \cite{Wolpaw2002,Nicolas-Alonso2012}, with user intent then determined through signal analysis, feature identification, extraction and classification. These BCI signal processing chains often target relatively static electrophysiological signal features common in adult EEG recordings, relying on their a priori predictability in spatial/temporal/spectral feature selection and determining data outliers. However, assuming the adult features may not be appropriate for analyzing populations with more variable signal features, like children \cite{Kinney-Lang2016}. 

Signal properties and profiles of the brain are continually changing from birth through adulthood \cite{Kinney-Lang2016,Marshall2002,Gasser1988}. For example, the location and frequency of the well established EEG alpha rhythm in adults is thought to migrate throughout childhood from approximately 6-9 Hz until reaching 8-13 in adulthood \cite{Gasser1988,Matsuura1985,Miskovic2015}. EEG signal recordings from young children are further confounded by high background noise alongside the shifting EEG signal bands \cite{Marshall2002,Matsuura1985,Miskovic2015}, resulting in obfuscated and weaker signals of interest. Therefore a means to identify and extract EEG features sensitive to the changing developmental profiles of children would be a critical tool in constructing paediatric BCI rehabilitation paradigms.

Tensor (or multi-way) analysis \cite{Cichocki2015,Acar2009} provides a potential framework to capture the dynamic developmental profiles in paediatric EEG, through investigating the relationships present in the multi-dimensional EEG data \cite{Phan2010,Cichocki2008}. Tensor analysis is a higher-order (i.e. multi-dimensional/multi-way) extension of standard matrix analysis techniques, which retains informative structural relationships between dimensions (domains or ways) in the data \cite{Cichocki2008}. Tensor analysis has already been adapted to adult BCI paradigms \cite{Cichocki2008,Cichocki2015,Liu2014}, and thus offers a structure to build tools towards effective paediatric BCI paradigms. 

This paper is an extension to our previous conference submission \cite{Kinney-Lang2017} which provided a proof-of-concept for characterizing developmental feature profiles of children using tensor analysis on an EEG dataset. Here an extended, robust feature selection paradigm is presented, introducing tensor component selection and model validation. Furthermore, the improved paradigm is demonstrated on several resting-state paediatric EEG datasets which span 1.) a rapidly developing preschool population with potential developmental impairments; 2.) a population spanning childhood to adulthood with potential developmental impairments; 3.) a healthy population during a stable developmental period of childhood. Successful characterization of key age-specific features within each dataset supports this approach as a potential adaptive tool for development-sensitive feature selection in paediatric EEG for applications like BCI.

\section*{Materials and Methods}

\subsection*{Datasets}

\subsubsection*{Muir Maxwell Epilepsy Centre}
A retrospective analysis of an epileptic preschool cohort ($<5$ years) from the Muir Maxwell Epilepsy Centre was included in this study, henceforth referred to as the MMEC dataset. The original cohort was prospectively recruited from National Health Service (NHS) hospitals in Fife and Lothian as part of the NEUROPROFILES study \cite{M.B.2015}. A 32-channel, unipolar montage captured routine EEG in the standard 10-20 system for each child. Of 64 children available, 14 were excluded from this study due to corrupted EEG data, inconsistent or incompatible EEG acquisition parameters and irregular recordings, resulting in a dataset of routine EEG from $n=50$ preschool children. If multiple resting-state EEG recordings existed, only the first recording was selected for each child to avoid weighting results toward children with more recordings and to select from the same awake resting-state data across all children.

\subsubsection*{Children's Hospital Boston-MIT} 
Publicly available data from a study at the Children's Hospital Boston-MIT of epileptic patients from infancy to early adulthood \cite{Shoeb2009a} was used in this study, downloaded through Physiobank.org \cite{Goldberger2000} and henceforth referred to as the CHB-MIT dataset. A 28-channel, bipolar montage captured EEG recordings continuously over two days of monitoring. Of the 23 patients available, 6 subjects were determined to have inconsistent EEG acquisition parameters for this study, due to discrepancies in the montages and unsuitable recordings. This resulted in $n=17$ subjects (age 2-19 y.o.) for analysis. The 48-hour continuous recordings were separated into 4-hour time blocks, with an equal number of trials at each time selected for processing. Results were averaged across all time bins to render a holistic representation of the resting-state data for each subject.

\subsubsection*{Child Mind Institute}
Resting-state EEG data for healthy control participants was taken from the open science resource provided by \cite{Langer2017} and the Child Mind Institute. Data captured from high-density 129-channel resting-state paradigms of pre-adolescent subjects (one age 6, the rest age 8-11 y.o.) was used. Of 45 subjects available, one subject was excluded (age 11) due to abnormalities in EEG processing, resulting in $n=44$ subjects for analysis. The single 6-year-old in the dataset was grouped with the 8-year-old class to allow for cross-validated classification.

A summary of subject distribution per age for each dataset is included in Table \ref{DatasetAgedist} of the supplementary data.

\subsection*{Pre-processing}
Raw EEG data was processed using the Fieldtrip toolbox \cite{Oostenveld2011} in Matlab 2015a. A two-pass (zero-phase forward and reverse) bandpass filter between [0.5-31] Hz was applied to EEG time-series signals. The filter was detrended, and signals were separated into 10-second (5-second) long trials for the MMEC/CHB-MIT (CMI) datasets. EEG channels were re-referenced to a common average reference and auxiliary/reference specific channels were removed. The EEG channel montages in the MMEC and CHB-MIT were matched through adapting the bipolar EEG electrode information from the CHB-MIT to the unipolar MMEC setup. The high-density set-up of the CMI data was not adapted to this same convention to avoid potential information loss. Channels in the CMI data with NAN values were removed. Trials with any seizure activity were immediately excluded from processing. A multi-pass artifact rejection system removed muscle, jump and ocular artifacts automatically, followed by manual inspection of data to verify and remove any remaining artifacts. Automatic rejection was based on recommended thresholding values given in Fieldtrip.

\subsection*{Tensor Construction}
Three-way tensors consisting of $[Spatial] \times [Spectral] \times [Subject]$ dimensions were created using the EEG channel, power spectra and subject age data for each dataset, resulting in $(19) \times (301) \times (50)$, $(19) \times (301) \times (17)$ and $(105) \times (61) \times (44)$ elements for the MMEC, CHB-MIT and CMI datasets respectively. Figure \ref{Figure1TensorConstruct}(a) provides a general illustration of the tensor construction. 

Time-frequency analysis of clean trials using Fieldtrip's multitaper method with a 0.5s Hanning window provided power spectra for each subject in the MMEC and CHB-MIT at 0.1 Hz resolution, and 0.5 Hz resolution for the CMI with data normalized. The power spectra in each subject was averaged across all trials providing a general spectral profile of the resting-state EEG for the $[Spectral]$ domain. 

The $[Subject]$ domain in each dataset was specifically ordered so all subjects were strictly increasing in age from youngest to oldest. The structure of this domain is critical as a developmental proxy, through which features can be identified as age-specific or pan-developmental as indicated by their general association to age in years.

\subsection*{Tensor Factorization: Parallel Factor Analysis}
Tensor factorization (decomposition) is a multiway extension of standard matrix analysis techniques (e.g. principal/independent component analysis) which describes a model of the structural relationship between tensor modes \cite{Cichocki2015,Phan2010,Kolda2008}. A common tensor factorization model for EEG data is Parallel Factor Analysis (PARAFAC, also known as Canonical Polyadic Decomposition CANDECOMP) \cite{Harshman1970,Klaas2003,Cong2015}, which decomposes a tensor \textbf{\underline{X}} into a linear combination of rank-1 tensors coupled with a super-diagonal core \cite{Harshman1970,Klaas2003}. EEG data readily supports a PARAFAC model due to its inherent higher-order structure, e.g. relationship between the time-series, channels and power spectra \cite{Cong2015}. 
Equation (\ref{parafac_eq}) illustrates the general \textit{R}-component PARAFAC model of a 3-dimensional array $\textbf{\underline{X}}(I \times J \times K)$:
\begin{equation}
x_{ijk} = \sum_{r=1}^{R} a_{ir}b_{jr}c_{kr} + e_{ijk}
\label{parafac_eq}
\end{equation}

\noindent with $x_{ijk}, a_{ir}, b_{jr}, c_{kr}$ $(i=1...I; j=1...J,k=1...K)$ and $e_{ijk}$ as elements of $\textbf{\underline{X}}$, domains $\textbf{A}(I \times R)$, $\textbf{B}(J \times R)$, $\textbf{C}(K \times R)$ and residual $\textbf{\underline{E}}(I \times J \times K)$, respectively. Figure \ref{Figure1TensorConstruct} B) illustrates tensor decomposition for varying components $R$ for a 3-dimensional tensor.

Tensor datasets were factored using an adapted PARAFAC function from the NWAY-toolbox for Matlab \cite{Andersson2000}. Several domain constraints were used in analysis to improve interpretation of results, and account for domain-specific properties like strictly non-negative components in the power spectra. Non-negativity constraints were applied to the $[Spatial]$ and $[Spectral]$ domains, with unimodality applied to the $[Subject]$ domain. Unimodality was imposed in order to extract components that are bound to specific age groups within a dataset. The $[Subject]$ domain structure permitted unimodal constraints since no subjects had repeated ages (data analyzed at `months-old', grouped into `years-old' for classification).

The PARAFAC model decomposition guarantees a 1:1 interaction between extracted factors across domains due to its super-diagonal core \cite{Cichocki2008,Klaas2003}. Through imposing strict zero-values in the core tensor on all but the diagonal components (i.e. making the core super-diagonal), any given component in a domain can only interact with the corresponding component in other domains. In example, in Figure \ref{fullpageMMECtensor} the first factor (\textit{black}) in the $[Spatial]$ domain corresponds directly with only the first factor (\textit{black}) in the $[Subject]$ and $[Spectral]$ domains as a result. Therefore, examining component interactions across domains in PARAFAC provides direct insight into the structural relationships within the data. In the presented work, this amounts to information on the extracted developmental features present throughout childhood.

Exploiting the mild conditions required for uniqueness of the PARAFAC model guarantees that the low-rank factor matrices of the PARAFAC decomposition retain their meaning \cite{Cichocki2015,Kolda2008}. In this proposal, retaining decomposition uniqueness allows interpretation of how the developmental $[Subject]$ domain influences $[Spatial]/[Spectral]$ factors. Even if the underlying relationships are not obvious in the original EEG data, they are assured to be viable through the uniqueness condition. Generic uniqueness of PARAFAC holds under the sufficient condition \cite{Kolda2008}:
\begin{equation}
\sum_{n=1}^{N} k_{\textbf{A}^{(n)}} \ge 2R+(N-1)
\label{parafac_uniq}
\end{equation}
for an \textit{N}-way tensor with \textit{k} elements for each domain matrix $\textbf{A}^{(\textit{n})}$ (i.e. mode-$n$ rank of A) and $R$ factor components \cite{Kolda2008}.  

\subsection*{Tensor Factorization: Component Selection}
Component selection is a critical step in tensor factorization. Choosing the optimal number of component factors for PARAFAC decomposition balances model suitability with proper representation of latent structural information. The unknown underlying developmental profiles in the data tensors are best captured by an unknown number of components, e.g. some non-minimal rank $R$ decomposition. In example, assume that there is a known number of EEG `sources', $s$ underlying the power spectra of the $[Spectral]$ domain. Each source $s$ could be described by exactly one rank-1 tensor by selecting $r = s$ components for tensor decomposition. However, if there are more sources $s$ than components $r$ (i.e. too few components chosen) then the model may obscure the less obvious, but still important, structural relationships. Similarly, if too many components are chosen, this may lead to over-fitting the model. To help gauge viable component selection of a PARAFAC model, the core consistency diagnostic (corcondia) is used \cite{Bro2003}. Corcondia provides a direct measure of the suitability of a specific PARAFAC model in an easy to interpret fashion by describing the degree to which the approximate tensor model deviates from its `true' super-diagonal core \cite{Bro2003}. The validity of the model can then be determined, with models under 40\% corcondia considered as generally non-viable \cite{Bro2003}. 

Corcondia may also provide insight into the point in which latent structural relationships have been matched uniquely to underlying component `sources'. Corcondia has a property of permanently and sharply decreasing after some maximum number of components is chosen \cite{Bro2003}. This decrease is not necessarily monotonic, but rather assured to never be better than the corcondia given at said maximum number of components. This was considered to be the point in which the maximum number of possible underlying relationships have been taken into account, i.e. there is likely maximum matching between components and underlying sources. Therefore, the number of components immediately prior to the downfall can serve as an inflection point for determining viable versus non-viable models, as done in this study. 

\begin{figure}
\centering
\includegraphics[width = 8.5 cm]{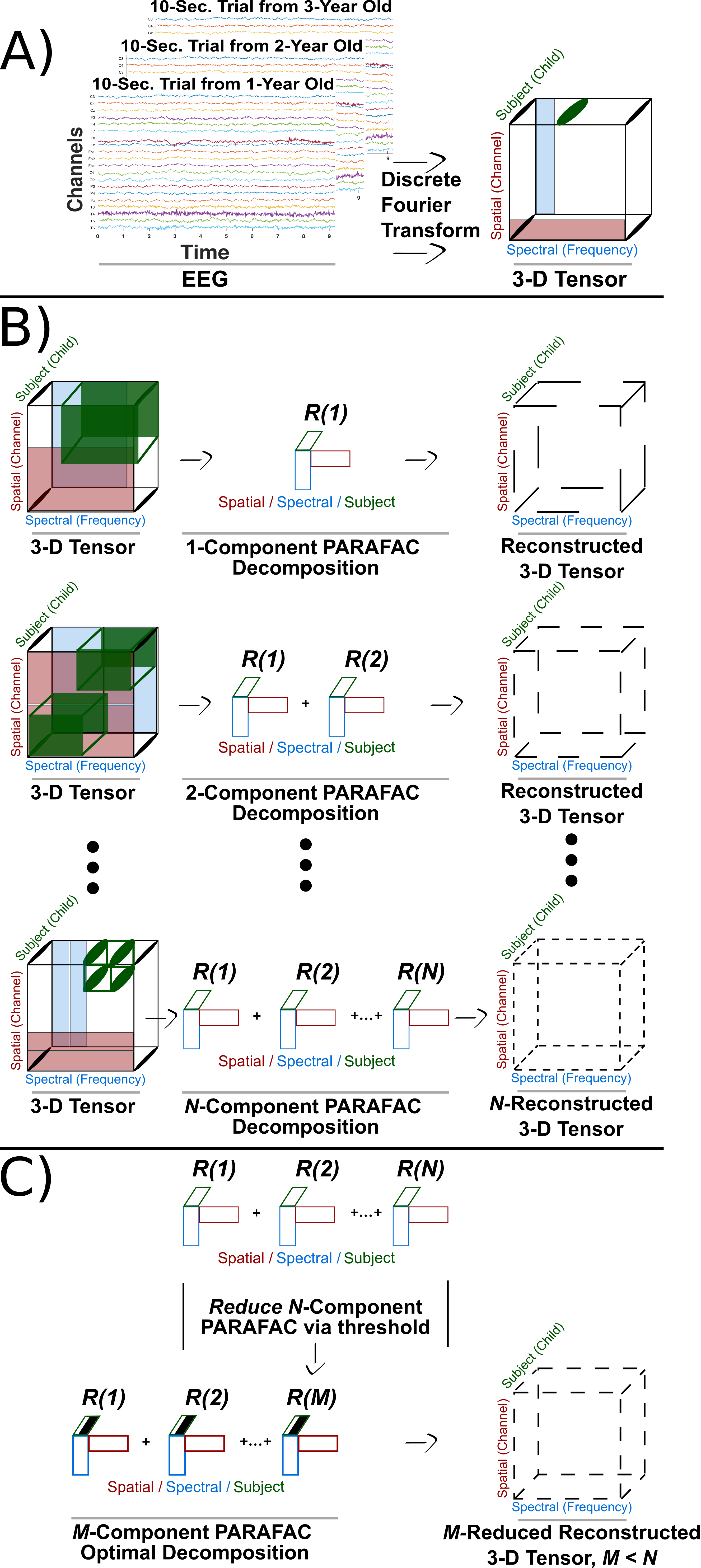}
\caption{ PARAFAC tensor construction and decomposition flowchart. A) Constructing a 3-dimensional tensor from raw resting-state time-series EEG with $[Spatial] \times [Spectral] \times [Subject]$ domains on the $X,Y,Z$ axes respectively. The $[Subject]$ domain is ordered to increase strictly with age. B) Illustration of $R=1:N$ component grid search, where each $R$-factorized model is used to reconstruct the original data to test corcondia and model suitability. C) Reduction of $R_N$-components to a optimal subset $R_M$-components (with $M < N$) based on a threshold $[Subject]$ domain and model validation via corcondia. Full resolution figure available upon request. }
\label{Figure1TensorConstruct}
\end{figure}

A two-step analysis chain over a varying set of components ($r = 1:20$) was constructed for this study to identify PARAFAC models which accounted for the maximum number of pan-developmental structures in the tensor while maintaining acceptable model viability. First, for each number of components $r$, five PARAFAC decompositions were run simultaneously with their corcondia and explained variance recorded. Using multiple simultaneous runs helped account for potential model convergence to local minimums. Any model with corcondia below 70\% was considered to be non-viable and was removed. The best fit model was considered to be the maximum rank decomposition model still above the 70\% corcondia. Then, a thresholding method based on our proof-of-concept work \cite{Kinney-Lang2017} was applied to select only the subset of factors which spanned the unimodal $[Subject]$ domain. Reduction of the factors to a development-specific subset helped remove any features reflecting properties of a single child, which could occur due to the unimodal constraint \cite{Kinney-Lang2017}. The reduced PARAFAC model then had its new corcondia evaluated, and the reduced model with the largest $R$-components maintaining corcondia above 70\% in both steps was selected for analysis and training. Figure \ref{Figure1TensorConstruct}(b) and Figure \ref{Figure1TensorConstruct}(c) outline the grid search, corcondia evaluation and optimal model selection process. Supplemental Figure \ref{corcondiaReducedCompare} shows full and reduced tensor model viability based on corcondia calculations for multiple components $r$ for one dataset.

\subsection*{Classification}
A modified direct projection of the extracted training factors onto the testing components was used for prediction validation, stemming from \cite{Escudero2015}. Direct projection traditionally includes a pseudo-inverse step, which inherently introduces negative testing component values \cite{Escudero2015}. Given the non-negative constraints on the training tensor decomposition, to retain meaningful prediction the projected test components must also be non-negative. Therefore, the non-negative least square (NNLS) solution of the Khatri-Rao product ($\odot$) between the non-$[Subject]$ dimensions (\ref{proj_eq1}) from the training tensor was introduced as an alternative to the pseudo-inverse step. Using the NNLS maintains the $[Spatial]$ and $[Spectral]$ non-negative domain constraints while still fulfilling the same approximate function as the pseudo-inverse. The NNLS solution was then multiplied by the $[Subject]$-domain matricized test tensor (\ref{proj_eq2}), resulting in a predicted $[Subject \times Factor]$ matrix for validation (\ref{proj_eq3}). Results from (\ref{proj_eq1}) were then multiplied by the $[Subject]$-domain matricized test tensor (\ref{proj_eq2}), resulting in a new predicted $[Subject \times Factor]$ non-negative matrix for validation classification.
\begin{equation}
ProjectedFactor_{train} = NNLS([Train_{spectral}]^{T}\odot[Train_{spatial}])
\label{proj_eq1}
\end{equation}
\begin{equation}
Test_{matrix} = [Test_{subject}]\times[Test_{spatial \cdot spectral}]
\label{proj_eq2} 
\end{equation}
\begin{equation}
Predicted_{test} = ProjectedFactor_{train} \times Test_{matrix}
\label{proj_eq3} 
\end{equation}  

To maintain stringent integrity for classification, data was split into training and testing cross-validation folds prior to tensor decomposition. A multi-class, ordinal classification scheme was devised to evaluate the tensor extracted factor's ability to predict subject age using the Weka toolbox \cite{Hall2009,Frank2016}. Subject age (in years) was used for within dataset class labels. An ordinal cost-matrix was used to account for the multi-class, ordinal nature of each data constructed tensor. The ordinal cost-matrix penalized misclassification through linearly weighted differences based on class age, thereby increasing classification penalties for predicting subjects as drastically older/younger compared to their actual age. The cost-matrix was matched for each dataset to the unique subject ages in that dataset.

Using the ordinal cost-matrix, a non-linear radial basis function (RBF) support vector machine (SVM) was trained using the decomposed PARAFAC factors for each cross-validation fold in each dataset. The RBF-SVM was optimized using a grid search in Weka to find $C$ and $\gamma$ which provided the highest classification accuracy. Results were evaluated on their overall classification accuracy and total penalty costs (e.g. the sum of all misclassification penalties based on the ordinal cost-matrix). Random classification and naive classification (e.g. choosing a single class for all subjects) is included for comparison. Results are reported as averages across all training folds with standard deviation and a two-tailed Student's \textit{t}-test to infer differences from random and naive classification.

The distribution of subjects per age in the MMEC and CMI datasets allowed for 4-fold and 5-fold stratified cross-validation respectively. The CMI classification included the single six-year-old as a member of the `Age 8' class to retain stratified cross-validation. Comparative classification within the CHB-MIT data was not possible due to the limited subjects per age.

\subsection*{Visualization}
To complement classification of the extracted high-dimensional tensor features across the $[Spatial],[Spectral],[Subject]$ domains, results from factorized training folds for the MMEC and CMI datasets are displayed using t-distributed Stochastic Neighbour Embedding (t-SNE) \cite{VanDerMaaten2008}. Using t-SNE, high-dimensional data can be visualized to capture both the local and global structure of the data through presence of clusters at several scales. Demonstrating t-SNE maps on individual training folds in the data offers a visual companion to the classification analysis, showing the local and global structure underlying a single fold used in analysis.

\subsection*{Simulations}
A simulation of pseudo-EEG data accompanies the real-world datasets. The Berlin Brain Connectivity Benchmark (BBCB) simulation code \cite{Haufe2016} was modified to include a shifting spectral frequency band of interest, similar to the alpha frequency seen in development \cite{Marshall2002,Matsuura1985,Gasser1988,Miskovic2015}. The band of interest lower bound was set with mean $\pm1$ standard deviation as the simulated age, up to 8 years-old. Afterwards, each additional simulated year increased the sampling mean thereby gating the lower bound towards 8-Hz. The upper bound was set at 3-Hz $\pm2$ standard deviations higher than the lower bound. The upper bound variation was at least 1-Hz above the lower bound. Ten children were simulated per age using the modified code, from 5 to 11 years-old. Simulated EEG was converted to Fieldtrip for processing, with a simulated tensor constructed in an identical fashion to the real-world datasets.

\section*{Results}
\begin{figure}
	\centering
	\includegraphics[width=\textwidth]{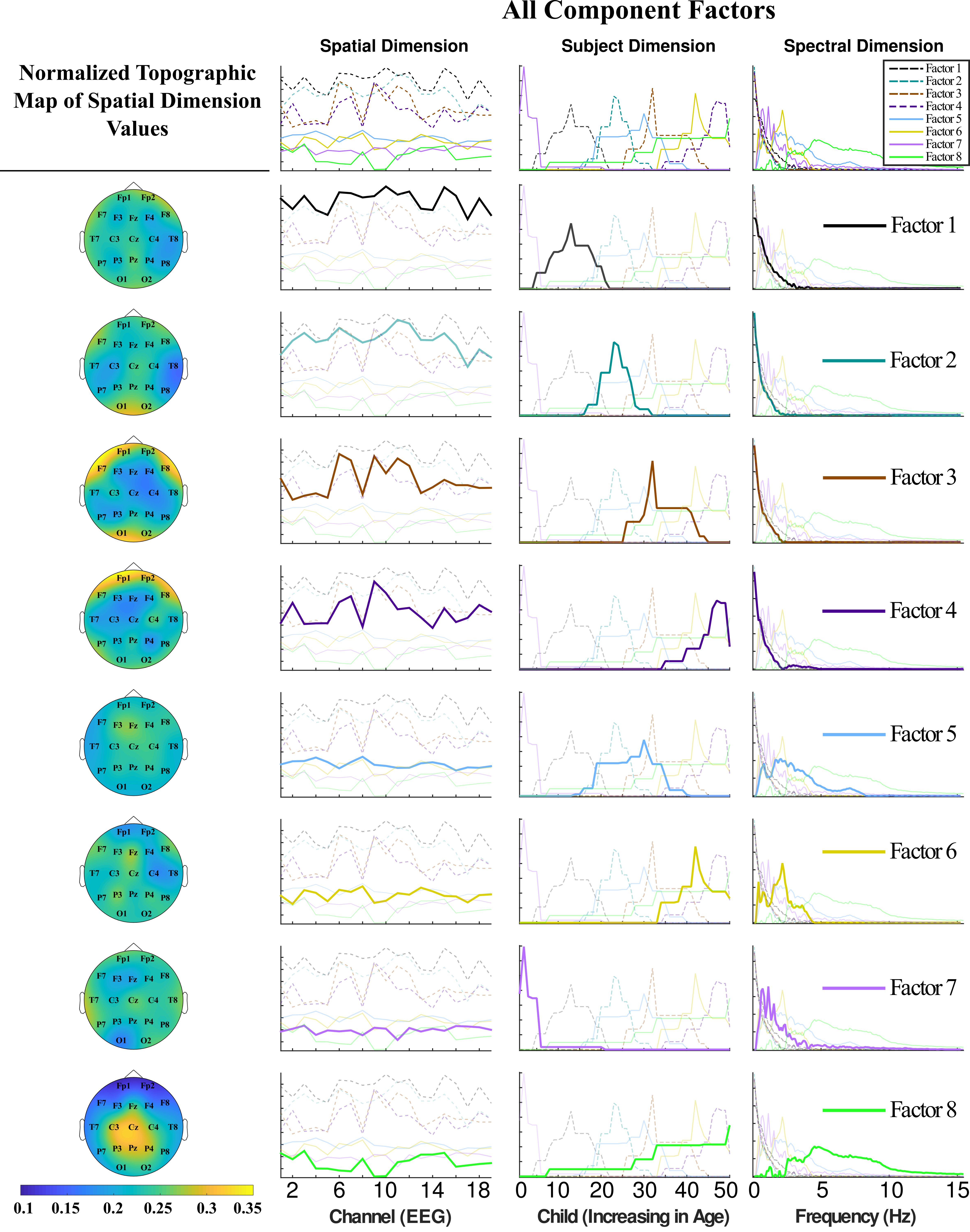}
	\caption{Detailed visualization of PARAFAC extracted developmental feature profiles of epileptic children from 0-5 years old from the MMEC dataset. Normalized topographic maps of the $[Spatial]$ domain in column 1 show EEG channel regions with higher/lower relative contribution for each individual feature. The $[Subject]$ domain x-axis is the child's number (e.g. Child 1, Child 2), organized by increasing age. The combined and separated feature profiles for the $[Spatial] \times [Subject] \times [Spectral]$ domains are shown in columns 2-4. Full resolution figure available upon request.}
	\label{fullpageMMECtensor}
\end{figure}

The proposed tensor analysis successfully identified latent developmental features across subjects independently for each dataset. A detailed visual breakdown of the PARAFAC model decomposition and its resulting `developmental profile' snapshots is given in Figure \ref{fullpageMMECtensor} as an example, using one entire dataset (MMEC). Qualitative developmental feature profiles are illustrated in each tensor domain, where actual weighted values in the testing folds were used for classification purposes. Individual factor contributions are shown in the extended profiles to help clarify the latent developmental relationships in each domain. Component profiles in the $[Subject]$ domain reflect in which ages the extracted factor (feature) is most dominant and influential. Features of the $[Subject]$ domain are ordered to match highest to lowest explained variance from the $[Spatial]$ domain. The normalized topographic map of the $[Spatial]$ domain shows relative regional contributions of EEG channels for each `developmental feature profile' (note this is not a topographic map of EEG activity).  The $[Spectral]$ domain is shown up to 15 Hz, as higher frequencies for preschool children in resting-state data has little activity of interest and remains fairly flat.
 
\subsection*{Preschool children with epilepsy: The Muir Maxwell Epilepsy Centre data} 

Classification was significantly improved using the identified developmental factor profiles as features in the MMEC dataset. Table \ref{MMEC_classif_table} contains the average classification results for the MMEC extracted features across all folds. The PARAFAC extracted features improved upon random classification by approximately 50\%, and reduced total penalty cost by 9.5 points or 37\% compared to naive classification (Student's \textit{t}-test, $p<0.05$). To put the reduced penalty cost in another context, if every test subject was misclassified by approximately 2 years using naive classification, PARAFAC reduced misclassification to approximately only 1.3 year for each test subject. Naive classification accuracy was improved upon using SVM and the extracted features by approximately 37\% as well (Student's \textit{t}-test, $p<0.01$). Average corcondia across training folds was $85.74\pm4.86$.
\begin{table}[tbh]
\centering
	\begin{tabular}{ccc}

		\hline
		Classification	&	 Penalty Cost 	&	Accuracy (\%)\\ \hline
		SVM				&	 16.0 $\pm4.1$ 	&	30.0 $\pm3.5$ \\
		Random			&	 --				&	20.1 $\pm0.7^{\star}$ \\
		Naive			&	25.5 $\pm1.7^{\star}$ &	22.0 $\pm3.6^{\star\star}$ \\
		\hline
	\end{tabular}
	\caption{Average classification results across all MMEC cross-validation folds for SVM, Random and Naive classification for epileptic children 0-5 years of age. No penalty costs available for random classification due to its random nature.
		\newline $^{\star}$ Indicates significant difference from SVM using student's \textit{t}-test at $p<0.05$.
		\newline $^{\star\star}$ Indicates significant difference from SVM using student's \textit{t}-test at $p<0.01$.}
	\label{MMEC_classif_table}
\end{table}

While only a third of subjects were on average correctly classified in the MMEC dataset, the significantly reduced cost penalties indicate a move toward reductions in gross age misclassification (e.g. classifying a subject age 0 as age 3, 4, or 5). This is critical, as improved misclassification penalties indicate mistakes trended more towards closely related ages (i.e. $\pm1$ or 2 years) more often. These improvements along with the factor profiles imply success in identifying developmentally important features of preschool children's EEG.

\subsection*{Child-to-Adult epilepsy spectrum: The CHB-MIT data}
Results of the CHB-MIT dataset were used to demonstrate the scalable nature of the proposed analysis across a broad age range, for subjects under otherwise similar conditions (e.g. epilepsy afflicted subjects with matched EEG montages). Due to the limited number of subjects for each age (only one to two - see Supplemental Table \ref{DatasetAgedist}), meaningful classification was not possible for the CHB-MIT dataset. Instead a qualitative illustration of the general developmental profile trends is presented in Figure \ref{halfpageCHBtensor}. The extracted factor profiles reflecting dominating influences at specific ages which match expected developmental patterns \cite{Gasser1988,Matsuura1985} indicate successful characterization of development sensitive features. These results reinforce the likely generalizable nature of the tensor analysis.

\subsubsection*{Profiles}
\begin{figure}
	\centering
	\includegraphics[width=\textwidth]{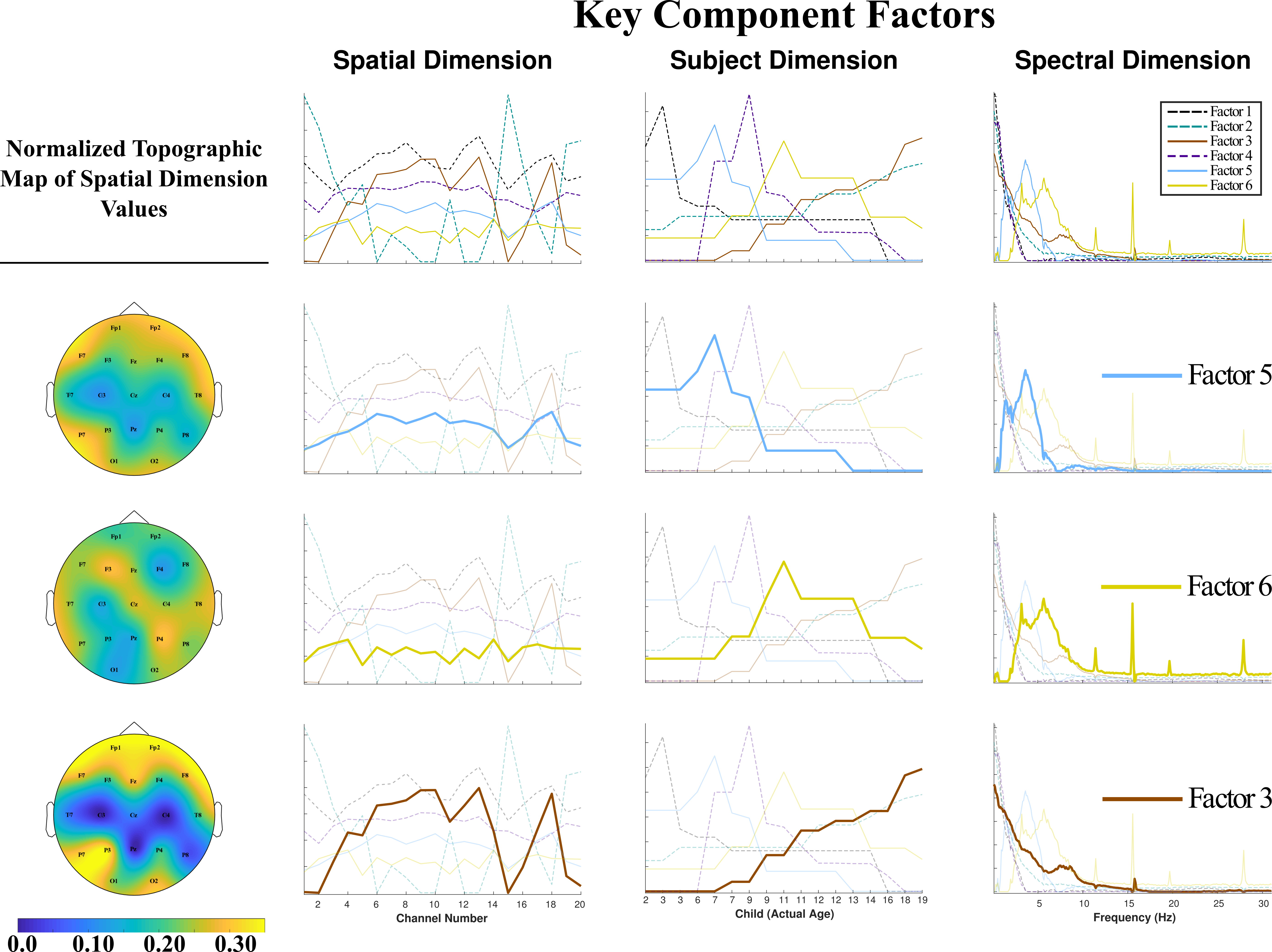}
	\caption{A compact visualization of the key feature profiles from the CHB-MIT dataset for epileptic subjects from age 2 to 19. A normalized topographic map reflecting relative contributions of each EEG channel region to the extracted factor is shown in the first column. Combined and separated feature profiles for the $[Spatial] \times [Subject] \times [Spectral]$ domains are shown in columns 2-4. Full resolution figure available upon request.}
	\label{halfpageCHBtensor}
\end{figure}

Figure \ref{halfpageCHBtensor} demonstrates the PARAFAC model decomposition of the CHB-MIT dataset in a condensed format. Key developmental feature profiles are emphasized across the extracted feature domains, with probable `background' profiles unaccented. The key feature profiles have been organized by the $[Subject]$ domain, with influential features prominent in early childhood to early adulthood ordered from top to bottom. The exact age of each subject is present on the $[Subject]$ domain axis. Sharp peaks in the $[Spectral]$ domain at 16, 19, 28 Hz are potentially residual artifacts from the time-frequency analysis and NAN-averaging across time-bins. Corcondia was 73.49 for the extracted factors, with 85\% explained variance.

\subsection*{School-age controls: The Child Mind Institute Data}
Developmental features were successfully identified for the healthy control CMI dataset via PARAFAC. While the available control data could not be age-matched to the MMEC for direct comparison, factor profiles found in the CHB-MIT data (seen in Figure \ref{halfpageCHBtensor}) at ages similar to CMI subjects supported analyzing the CMI dataset for feasibility in identifying developmentally sensitive features in normally developed children. Table \ref{CMI_classif_table} contains the average classification results across the training/testing folds for CMI subjects. PARAFAC extracted features improved significantly upon random classification by approximately 25\%, and reduced total penalty cost by 2 points or 19\% compared to naive classification. Average corcondia across training folds was $78.25\pm6.57$.

Although classification results had smaller improvements in classifier performance and penalty reduction compared to the epileptic population of the MMEC dataset, significant improvements were still present using the PARAFAC model for SVM classifier training. Extra difficulty in discerning developmental differences in the CMI data was to be expected, given the smaller developmental window and higher homogeneity of subjects in the CMI dataset. Therefore, despite only marginal boons in accuracy and penalty costs, the classification improvements in the CMI dataset under PARAFAC are important. The positive results serve as evidence that the proposed tensor analysis is accessible to `normal' developing children.

The CMI dataset results also further illustrates the scalable nature of the tensor analysis, demonstrating successful feature extraction on subjects more similar developmentally (i.e. healthy, between the ages of 8-11 y.o.) compared to the other datasets. Physiological changes across this age span are significantly less drastic compared to both the MMEC development window from infancy to early childhood (age 0-5 y.o.), and the CHB-MIT development window from infancy to early adult hood (age 2-19 y.o.). 

\begin{table}[tbh]
	\centering
	\begin{tabular}{ccc}
		
		\hline
		Classification	&	Penalty Cost 			&	Accuracy (\%) \\ \hline
		SVM				&	9.6 $\pm0.9$ 			&	34.2 $\pm1.9$ \\
		Random			&	-- 						&	27.4 $\pm1.3^{\star\star}$ \\
		Naive			&	11.8 $\pm1.1^{\star}$ 	&	29.8 $\pm3.6$\\
		\hline
	\end{tabular}
	\caption{Average classification results across all CMI cross-validation folds for SVM, Random and Naive classification between healthy children 8-11 years of age. No penalty costs are available for random classification due to its random nature. 
		\newline $^{\star}$ Indicates significant difference from SVM using student's \textit{t}-test at $p<0.05$.
		\newline $^{\star\star}$ Indicates significant difference from SVM using student's \textit{t}-test at $p<0.01$. }
	\label{CMI_classif_table}
\end{table}

\subsection*{t-SNE Visualization}
\begin{figure}
	\centering
	\includegraphics[width = 15 cm]{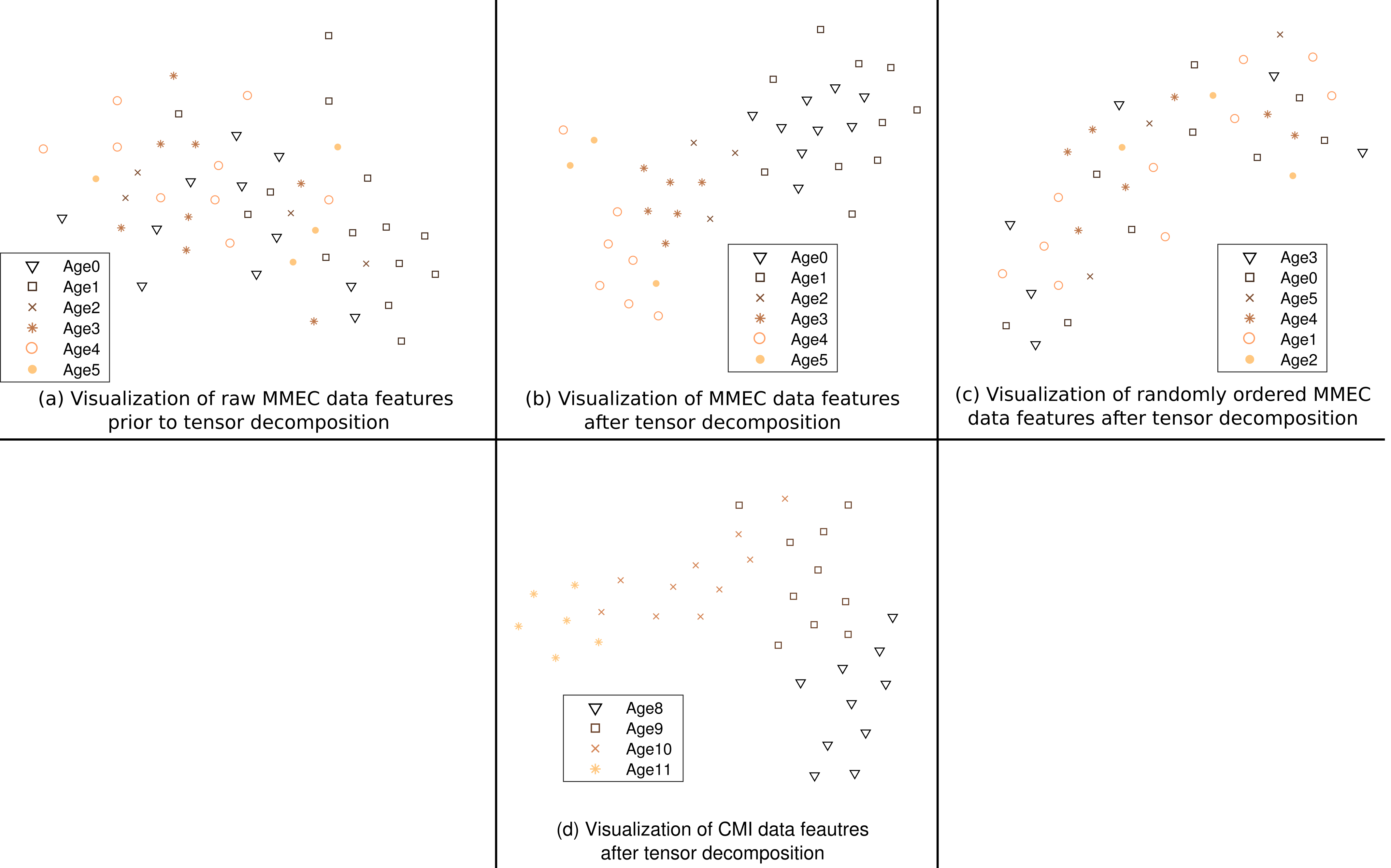}
	\caption{t-SNE Map visualizations with features grouped by subject age. (a) Map of EEG data prior to tensor decomposition of the MMEC dataset. (b) Map of PARAFAC extracted developmental features of the MMEC dataset, using the first cross-validaiton training fold. (c) Map of PARAFAC extracted developmental features of the MMEC dataset when the original tensor has a randomly ordered $[Subject]$ domain. (d) Map of PARAFAC extracted features from the CMI dataset using the first cross-validation training fold. Full resolution figure available upon request.}
	\label{tsneFig}
\end{figure}

\begin{figure}
	\centering
	\includegraphics[width = 8 cm]{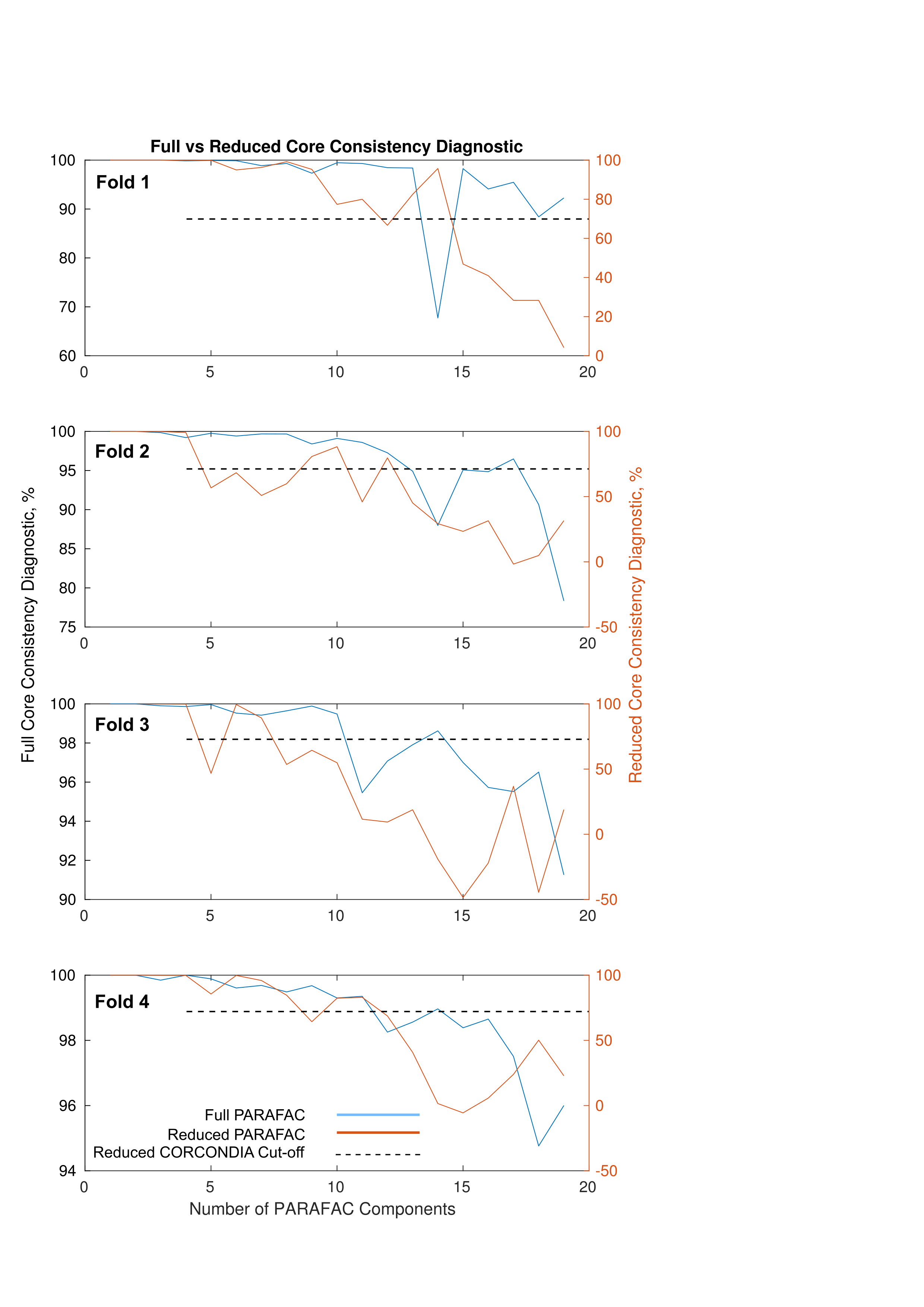}
	\caption{Full and threshold reduced PARAFAC model corcondia for each fold in a factorization of MMEC dataset. Only a decomposition maintaining 70\% corcondia for both models was be considered viable, with the reduced tensor used for classification. Full resolution figure available upon request.}
	\label{corcondiaReducedCompare}
\end{figure}

Features from the first training fold tensor decomposition of the MMEC and CMI datasets are displayed as t-SNE maps in Figure \ref{tsneFig}(b) and Figure \ref{tsneFig}(d) respectively. Both t-SNE maps demonstrate strong local grouping of different age groups. Maps of EEG data prior to tensor factorization (Figure \ref{tsneFig}(a)) and when the decomposition has a randomly ordered $[Subject]$ domain in the MMEC data (Figure \ref{tsneFig}(c)) are included for comparison. The results demonstrate significantly improved feature grouping and clusters using the properly ordered tensor decomposition methodology compared to using the raw EEG or random ordered $[Subject]$ domain. The t-SNE maps help illustrate how SVM classification from the raw EEG time-series data using all frequency and spatial features do not perform better than random classification in the MMEC. Similar patterned results were found for CMI data.

\subsection*{Simulation}
A 3-component PARAFAC model revealed the underlying `ground truth' developmental profiles built into the simulation tensor data. Resulting component factors are presented in Supplemental Figure \ref{SimulatedEEG}. Corcondia was 99\% for the model, with approximately 23\% explained variance. These values are understandable for the model, as the BBCB pseudo-EEG simulation was designed to retain trilinearity in the data while introducing multiple levels of noise at the frequency band, brain background and sensor layers. Replicating developmental profile extraction results in the simulated data grants further support to our conclusions in real-world datasets.

\section*{Discussion}
 The tensor analysis outlined in this paper lays a foundational framework capable of extracting latent structures and features associated with development in paediatric EEG. The unsupervised nature of this framework opens the door to encourage better personalized paradigms for data-driven technologies aimed at paediatric populations. Capitalizing on these developments could help translate new technologies to children, which are sensitive to developmental features in EEG.

\subsection*{Tensor analysis derives informative `development feature profiles' of paediatric subjects}
The factor profiles derived in this study confirm the proposed PARAFAC decomposition simultaneously accounts for background EEG noise and shifting frequency bands across subject age, often explaining more than 85\% of the data variance. Low-frequency, high power spectral activity typically associated with background EEG noise can be seen in Figure \ref{fullpageMMECtensor}, with factors 1-4 reflecting decaying power curves in the $[Spectral]$ domain, coupled with relatively strong contributions across all channels in the $[Spatial]$ domain. These factors are likely characterizing general `background' EEG noise independent to potential signals of interest (e.g. factors 5-8). Critically, the $[Subject]$ domain demonstrates the tensor analysis has sensitivity to subtle developmental differences, since the features uniquely correspond to distinct age groupings, even within the potential background noise. 

Shifts in spectral power and frequency due to development \cite{Marshall2002,Matsuura1985,Miskovic2015} are seen in the qualitative factor profiles of both the MMEC and CHB-MIT datasets. In the MMEC dataset, for example, the very low frequency, high power dominated spectral profile associated with infant and early-life EEG recordings seem to be reflected by Factor 7 in the $[Subject]$ and $[Spectral]$ domains. Meanwhile, factor 5 (light blue) is centrally located in the $[Subject]$ domain (covering approximately ages 1-3) and spans the 3-7 Hz range of the $[Spectral]$ domain. Factor 5 therefore highlights the likely dominant frequency range for those ages, while also reflecting a shift in power towards higher frequencies, which is expected with growing. Factor 8 illustrates a further shift towards higher frequencies contributing more to the spectral profile, as it steadily increases for subjects 31-50 (approximately ages 3-5 y.o.) in the $[Subject]$ domain. Also, factors 6 and 8 may represent the beginning separation between classical EEG bands of interest, i.e. the delta/theta bands and the alpha/beta bands respectively.  

Similar developmental shifts are also seen in the CHB qualitative feature profiles. The key component factors show significantly reduced spectral power contributing more in higher $[Subject]$ domain ages, alongside a shift in the $[Spectral]$ dimension towards higher frequencies. These extracted profiles are reflective of the traditional movement and prevalence of the classical EEG bands, like alpha, throughout child development into adulthood \cite{Matsuura1985,Miskovic2015}. 

\subsection*{Improved classification results verify tensor extracted features' sensitivity to development}
Improvements in classification coupled with the obtained tensor profiles of both impaired (epileptic) and healthy children indicate age-specific factors uncovered from EEG via PARAFAC contain structural information on latent developmental relationships. The scalable nature of the proposed analysis showed promise in identifying relevant features to development across varying developmental conditions, including both afflicted/healthy populations, and at slow/rapid developmental windows. 

With better characterization of such features for the paediatric subjects comes a stronger case for translating signal processing and machine learning applications to children. Clear support for this is seen in the comparison between t-SNE maps for the full feature raw EEG time-series data and the PARAFAC processed data. Using the feature-full raw EEG time-series without processing led to complete failure in identifying developmental features. The resulting t-SNE map has no discernible structure with random clusters and groupings. Characterization of the underlying developmental profile was rendered completely imperceptible, which is likely reflected in the classification being no better than chance. On the other hand, the highly-structured t-SNE maps of the tensor extracted features for both the MMEC and CMI dataset reflect well characterized developmental profiles, which can be utilized in machine learning applications.

Importantly, the randomly ordered $[Subject]$ domain t-SNE map suggests that successful identification of key developmental features is not inherent to tensor factorization itself. Although the overarching global structure is capable of being identified, as seen in the similarity of global shapes of Figure \ref{tsneFig}(b) and Figure\ref{tsneFig}(c)), the local grouping is completely lost. Rather, the viability of determining developmentally sensitive features relies on proper construction of the $[Subject]$ domain e.g. having an inherent proxy to growing like strictly increasing subject age. Future work could investigate the effects of altering the inherent structure of the $[Subject]$ domain to reflect other developmental markers, such as cognitive or behavioural scores. 

Exploiting the structural information from higher-order tensors constructed with careful construction of the $[Subject]$ domain as a proxy measure for child development provides a methodological framework designed to enhance sensitivity to physiological changes common throughout childhood. Higher sensitivity to these unique developmental profiles could be useful in applications like BCI. Through using a framework built to determine a current child's developmental state at an electrophysiological level, the BCI could be automatically tuned and weighted appropriately for subjects at different points along development. Additionally, the results could provide `healthy development curves' in studies for comparison to potentially developmentally impaired children.

Together these findings expand upon our previous results \cite{Kinney-Lang2017}. Our improved methodology is verified using multiple datasets. Results indicate the new proposed methods can account for developmental differences in background EEG and shifting spectral signals for children under a variety of different developmental conditions. The classification results illustrate a means for developmental feature extraction sensitive to progressive changes, while the profiles provide informative context regarding the relationship between $[Spatial]$ and $[Spectral]$ structures relative to subject age and development.    
 
\subsection*{Limitations}
Limitations in this study included restricted access to age/acquisition-matched paediatric datasets and the heterogeneity associated with epilepsy in subjects. Due to limited resources no direct age/acquisition-matched healthy control data was available for analysis to compare directly. To mitigate these drawbacks, however, multiple publicly available datasets were used to demonstrate the proposed methodology in multiple settings. The CHB-MIT dataset built upon our analysis from the MMEC across a wider age range using a similar disease condition, while the CMI dataset represented healthy control within the bounds of childhood (but not age-matched). Future work using a more homogeneous population with age-matched controls could help further validate the results, along with data from both resting-state and event-related EEG.

\section*{Conclusion}
Advanced signal processing, like PARAFAC, combined with machine learning can help distinguish non-obvious developmental patterns from child EEG data. This study proposes tensor analysis can provide an intuitive sense of the latent developmental relationships in paediatric EEG data, and provide a way for development-sensitive feature extraction. The results indicate successful identification of factor profiles and benefits to classification analysis for a wide variety of developmental conditions, including both afflicted and healthy paediatric populations. This study lays a methodological framework which could improve applications for children reliant on EEG processing and analysis, like BCI. Further development on this framework could help improve BCI application sensitivity to developmental changes by setting the groundwork for a `developmental domain' for tensor-based EEG classification in BCIs (e.g. \cite{Zink2016}). These advances could help immensely in translating the BCI technology to paediatric populations and pave the way for development of more readily accessible, effective rehabilitation strategies internationally.   
\section*{Acknowledgement}
Funding support for this project was provided by the RS McDonald Trust, Thomas Theodore Scott Ingram Memorial Fund and the Muir Maxwell Trust. The authors would like to thank the children, parents, clinicians and researchers who volunteered and comprised each of the datasets, and the Muir Maxwell Epilepsy Centre for providing access the MMEC dataset. The authors also thank Ephrem Zewdie for his comments and suggestions on figures.
\section*{References}
\bibliographystyle{unsrt}
\bibliography{mylibrary}

\section*{Supplementary Information}
\begin{table}[htb]
	\centering
	\begin{tabular}{|cc|}
		\hline
		\multicolumn{2}{|c|}{MMEC Dataset} \\ \hline
		\# of Subj. 	& 	Age (Months) \\
		11	&	0-12 \\	
		14	&	12-24\\
		4	&	24-36\\
		8	&	36-48\\
		10	&	48-60\\
		3	&	60-72\\ \hline\hline
		\multicolumn{2}{|c|}{CHB-MIT Dataset} \\ \hline
		\# of Subj. 	& 	Age (Years) \\
		1	&	2 \\	
		2	&	3 \\
		1	&	6 \\
		2	&	7 \\
		2	&	9 \\
		2	&	11 \\	
		2	&	12 \\
		1	&	13 \\
		1	&	14 \\
		1	&	16 \\
		1	&	18 \\
		1	&	19 \\ \hline\hline
		\multicolumn{2}{|c|}{CMI Dataset} \\ \hline
		\# of Subj. 	& 	Age (Years) \\
		1$^*$	&	6 \\	
		11	&	8 \\
		12	&	9 \\
		13	&	10 \\
		7	&	11 \\ \hline
		\multicolumn{2}{|c|}{\tiny{$^*$Included as 8 y.o. for classification}} \\ \hline
		
		\end{tabular}
		\caption{Distribution of subjects per age for each dataset.}
		\label{DatasetAgedist}
\end{table}

\begin{figure}
	\centering
	\includegraphics[width = \linewidth]{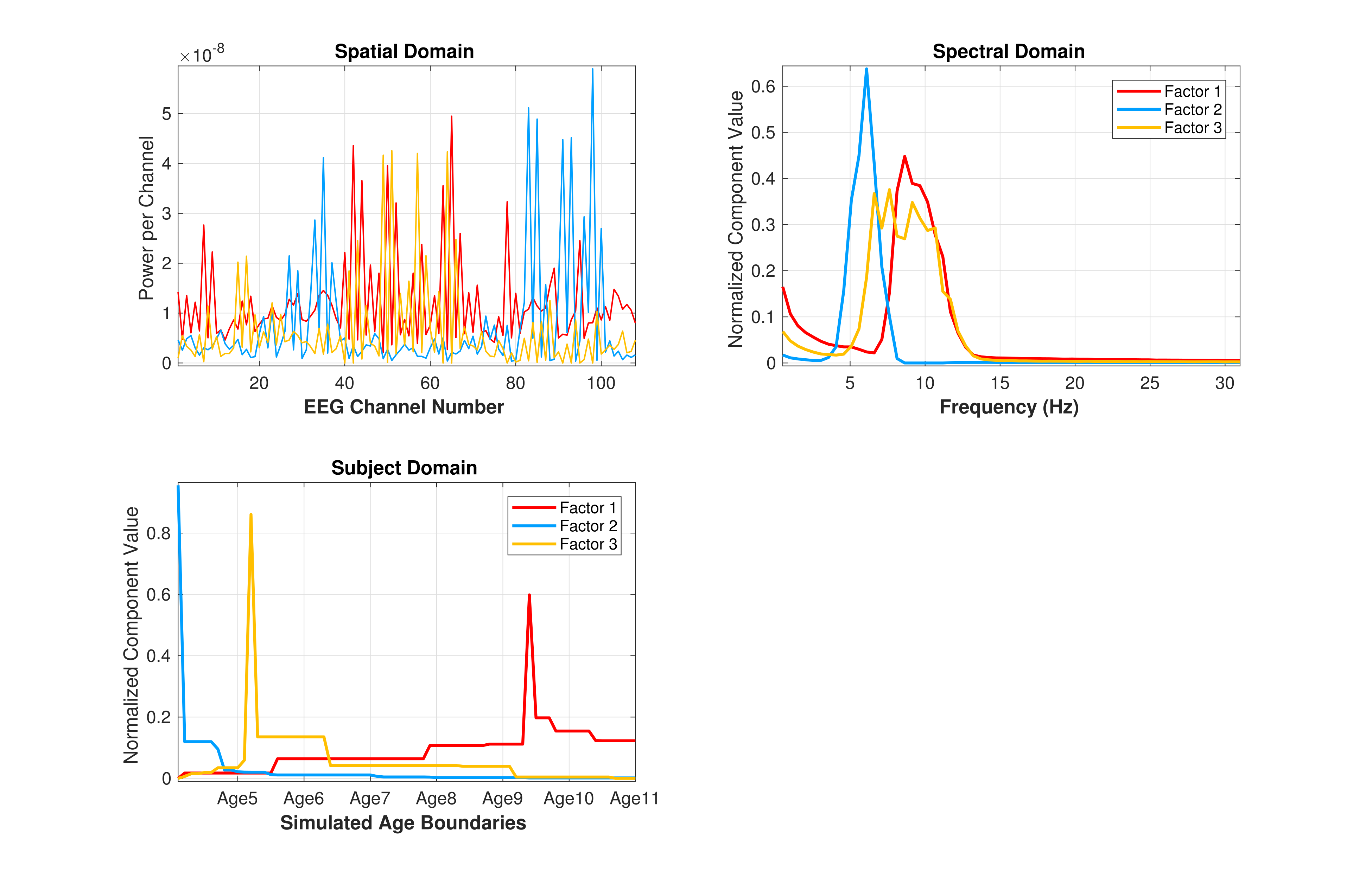}
	\caption{The 3-component PARAFAC model decomposition of the simulated tensor data. Raw power values are shown for the $[Spatial]$ domain, while component values in the $[Spectral]$ and $[Subject]$ dimension have been normalized. The 3-component model accurately identifies the underlying shifting spectral frequencies set to vary from approximately 5-8 Hz to 8-12 Hz based on age in the $[Subject]$ domain. Sharp peaks in the $[Subject]$ domain reflect the simulated child with the best signal-to-noise ratio. Full resolution figure available upon request.}
	\label{SimulatedEEG}
\end{figure}

\end{document}